\documentclass[prb,showpacs,superscriptaddress,twocolumn]{revtex4-1}

\usepackage{graphicx}  
\usepackage{dcolumn}   
\usepackage{bm}        
\usepackage{amssymb}   
\usepackage{multirow}

\usepackage{natbib}
\usepackage[final=true]{hyperref}
\usepackage{amsmath}
\usepackage{epstopdf} 

\usepackage[utf8]{inputenc}
\usepackage[T2A]{fontenc}

\usepackage{color}
\usepackage[usenames,dvipsnames]{xcolor}

\begin{document}

\title{Ratchet effect in graphene with trigonal clusters}

\author{S.V.~Koniakhin}
\email{kon@mail.ioffe.ru}
\affiliation{Ioffe Physical-Technical Institute of the Russian Academy of Sciences, 194021 St.~Petersburg, Russia}
\affiliation{St. Petersburg Academic University - Nanotechnology Research and Education Centre of the Russian Academy of Sciences, 194021 St. Petersburg, Russia}

\date{\today}

\begin{abstract}

In this study, we explore the ratchet effect in graphene with artificial, triangular scatterers from a theoretical standpoint. It is demonstrated that the skew scattering of carriers by such coherently oriented defects results in the ratchet effect in graphene, i.e., in a direct current under the action of an oscillating electric field. Scattering on various types of defects exhibiting threefold symmetry is considered in this paper: scattering on a cluster in the shape of a solid triangle in the classical and quantum mechanical limits, and scattering on three- point defects placed at the corners of a triangle. The DC current is calculated for a classical range of oscillating field frequencies.
\end{abstract}

\pacs{81.05.ue, 72.80.Vp, 72.10.Fk, 03.65.Nk}

\keywords{graphene, skew scattering, ratchet effect, photogalvanic effect}

\maketitle

\section{Introduction}

The physics of graphene has been most captivating topic in condensed-matter physics over the past decade. Particularly promising are the effects of nonlinear transport in graphene\cite{Glazov,0295-5075-79-2-27002,APL}, including the photogalvanic effect, a subclass of the ratchet effect \cite{epj1,Ivch,chep3,PhysRevB.86.115301} in which the appearance of a directed particle flux is driven by an external stochastic or periodic field. The flux is nonlinear with respect to the magnitude of the driving force.

The studies of ratchet effects are important both for the fundamental physics, because such effects are deeply related with the issues of second law of thermodynamics \cite{RevModPhys.81.387}, and for the future device applications, such as ratchet effect-based terahertz radiation detectors. The possibility of manufacturing of such devices was demonstrated for semiconductor two-dimensional structures\cite{PhysRevB.77.245304,PhysRevB.83.165320}.

This paper develops the consistent analytical and numerical theory of the ratchet effect in graphene. We consider the new type of symmetry breaking that leads to the ratchet effect in graphene, namely the threefold symmetry. This paper microscopically describes the skew scattering on triangular defects employing the second order quantum mechanical perturbation theory for deriving the numerical estimations of the ratchet current magnitude for the electric fields of classical magnitudes and graphene samples with reachable parameters. The obtained numerical values can be directly compared with future experimental data.

\begin{figure}
\includegraphics[width=1.0\linewidth]{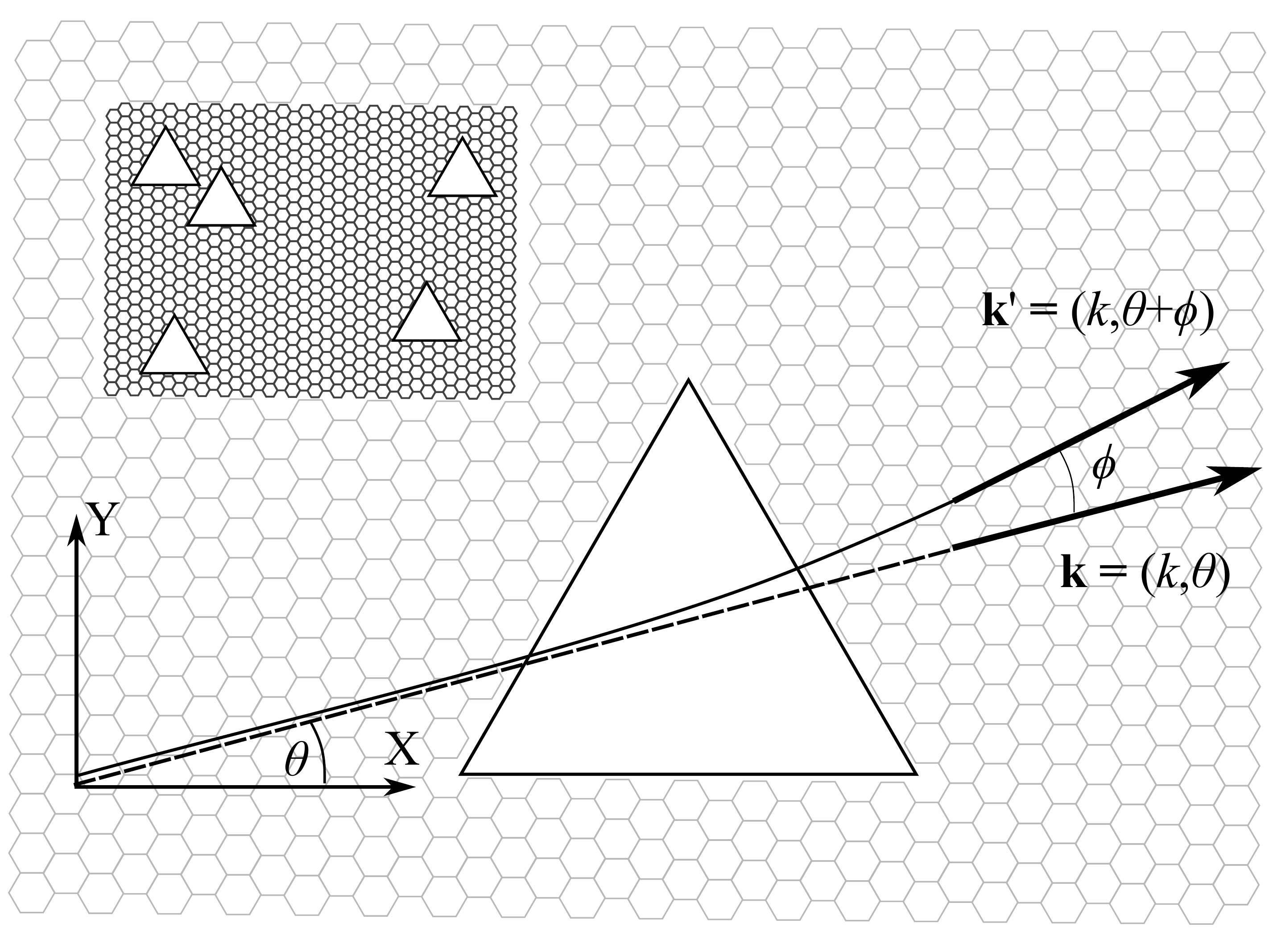}
\caption{Geometry of the problem. One of the triangle's sides is parallel to the X axis. The initial electron state, denoted as $\mathbf{k}$, has a polar angle $\theta$. The final state, $\mathbf{k}'$, has a polar angle $\theta'=\theta+\phi$. The inset shows a graphene sheet with an array of coherently oriented triangular clusters.}
\label{fig_Geometry}
\end{figure}

We start with phenomenological analysis of the effect based on the symmetry considerations because it allows one to predict the polarization and intensity dependence of the photocurrent without knowing the microscopic process of the current formation. Due to the considered shape of defects the system obeys $C_{3v}$ symmetry, and ratchet current in such system is given by\cite{PhysRevB.77.245304}
\begin{equation} \label{symmetry}
\left(\begin{array}{c}
j_x \\ j_y
\end{array}\right)= \chi \left(\begin{array}{c}
 E_x E_y^*+ E_y E_x^* \\ |E_x|^2-|E_y|^2
\end{array}\right).
\end{equation}
It can be observed that the transverse ratchet effect is allowed by the considered symmetry. These relations also describe the photogalvanic effect with normal incidence of radiation. Fig. \ref{fig_Geometry} shows the geometry of the problem.

Similarly to this work, ref. \cite{chep3} employed the Boltzmann kinetic equation approach to study scattering on semidisks, and ref. \cite{PhysRevB.86.115301} investigated ratchet effects in graphene with a noncentrosymmetric lateral potential. We provide calculations of the asymmetric part of scattering rate for several types of trigonal scatterers: a point defect trimer (PD trimer) and a cluster in the shape of a solid triangle. We study the case of the solid triangle using classical and quantum mechanical rules of elastic scattering. Finally, we implement the obtained asymmetric scattering rate in the Boltzmann kinetic equation and provide values of the ratchet current for structures that can be manufactured. De facto we derive the coefficient $\chi$ in relation \eqref{symmetry}.

Most part of previous works considering the skew scattering on clusters simulated the behavior of the Langevin particles \cite{PhysRevB.71.214303,PhysRevB.75.020201,1367-2630-11-7-073046,PhysRevE.78.041127,chep2}. Some other works studied the electron transport through the threefold quantum billiards \cite{0295-5075-44-3-341,PhysRevB.61.15914} and quantum--dot arrays \cite{QD}.

The following assumptions about the electronic structure of graphene are made. We study a degenerate gas of carriers and neglect the effects of trigonal warping \cite{PhysRevB.66.035412}, taking the linear isotropic dispersion $\varepsilon(\mathbf{k})$ to be equal to $\hbar v_F k$. Finally, we do not consider the effects of spin swap or intervalley scattering.
\subsection{Asymmetric scattering}
The origin of the ratchet effect is the breakdown of central inversion symmetry, which at the microscopic level can give rise to skew scattering (or asymmetric scattering). The most general property of the elastic scattering rate $W(\mathbf{k},\mathbf{k}')$ between the electronic states $\mathbf{k}$ and $\mathbf{k}'$ is the time invariance of the scattering rate:
\begin{equation*} \label{WasymmFinal}
W(\mathbf{k},\mathbf{k}')=W(-\mathbf{k}',-\mathbf{k}).
\end{equation*}
For a potential with central inversion symmetry, the more stringent condition is
\begin{equation*} \label{WasymmFinal}
W(\mathbf{k},\mathbf{k}')=W(-\mathbf{k},-\mathbf{k}')=W(\mathbf{k}',\mathbf{k}),
\end{equation*}
known as detailed balance. A scattering potential lacking central inversion symmetry leads to the breakdown of the detailed balance and, as a consequence, to the possibility of various intriguing phenomena.

Other origins of the photogalvanic effect and the anomalous Hall effect \cite{PhysRevLett.99.066604,PhysRevB.76.235312,PhysRevB.82.115124}  are side jumps and the effects of the Berry phase\cite{RevModPhys.82.1539}.

The standard collision integral can be divided into symmetric and  antisymmetric parts as follows:
\begin{multline*} \label{colliding_general}
St(f(\mathbf{k}')) =
\int_{BZ} d\mathbf{k} W^{symm}(\mathbf{k},\mathbf{k}')\left[f(\mathbf{k})-f(\mathbf{k}')\right]\\+
\int_{BZ} d\mathbf{k} W^{as}(\mathbf{k},\mathbf{k}')\left[f(\mathbf{k})+f(\mathbf{k}')\right],
\end{multline*}
where the symmetric and the asymmetric scattering rates are defined as
\begin{equation*}
W^{symm}(\mathbf{k},\mathbf{k}')=0.5 \left[W(\mathbf{k},\mathbf{k}')+W(\mathbf{k}',\mathbf{k})\right]\approx W(\mathbf{k},\mathbf{k}'),
\end{equation*}
\begin{equation*}
W^{as}(\mathbf{k},\mathbf{k}')=0.5 \left[W(\mathbf{k},\mathbf{k}')-W(\mathbf{k}',\mathbf{k})\right] \ll W(\mathbf{k},\mathbf{k}').
\end{equation*}
The designation $BZ$ implies integration over the entire Brillouin zone. This designation is formal and de facto only electronic states near the Fermi level are involved in the integration.

We can explicitly use the elasticity of scattering and rewrite the normalized asymmetric scattering rate using the absolute value of the electron wave vector $k$, polar angle $\theta$ of the initial electron wave vector $\mathbf{k}$ and the scattering angle $\phi$, which obeys the relation $\theta'=\theta+\phi$. Fig. \ref{fig_Geometry} shows the scattering geometry. Therefore, in spite of $W^{as}(\mathbf{k},\mathbf{k}')$, one may write  $W^{as}(k,\theta,\phi)$.

The symmetric part of the collision integral can be treated in the relaxation time approximation, and the asymmetric part provides a basis for introducing the operator of asymmetric scattering:
\begin{multline} \label{StAsymmIntro}
\hat{A} f(k,\theta) = \frac{1}{4\pi^2} \int kdk
\int_{-\pi}^{+\pi} d\phi \delta(k-k')\\ \times
W^{as}(k,\theta,\phi) \left( f(k,\theta)+f(k,\theta+\phi) \right).
\end{multline}

Fermi’s golden rule, which yields the transition rate for two-dimensional Dirac fermions, reads as follows:
\begin{equation*} \label{FermiGolden}
W(\mathbf{k},\mathbf{k}') d\mathbf{k}'=\frac{2\pi}{\hbar} |F|^2 \frac{S}{\hbar v_F} \delta(k-k') d\mathbf{k}'.
\end{equation*}
In the first approximation of perturbation theory, the scattering amplitude $F$ directly matches the scattering potential matrix element:  $F=V(\mathbf{k},\mathbf{k}')$. It can be immediately observed that the detailed balance is always met in this approximation. Taking into account the second-order approximation for the scattering amplitude  $F=F^{(1)}+F^{(2)}$, where
\begin{equation*} \label{F2ndorder_energy}
F^{(2)}(\mathbf{k},\mathbf{k}')=\frac{S}{4\pi^2}\int_{BZ} d\mathbf{q}\frac{V(\mathbf{k}'-\mathbf{q}) V(\mathbf{q}-\mathbf{k})}{\hbar v_F k - \hbar v_F q + i\delta},
\end{equation*}
will lead \cite{Belinicher:1980} to the following expression for the asymmetric part of the scattering rate
\begin{multline} \label{Wasymm1}
W^{as}(\mathbf{k},\mathbf{k}')d\mathbf{k}' = d\mathbf{k}'  \frac{S^2}{2\pi \hbar^3 v_F^2} \delta(k-k') \\ \times
\Im \left[ V(\mathbf{k},\mathbf{k}') \int_{BZ} d\mathbf{k}'' V(\mathbf{k}',\mathbf{k}'')V(\mathbf{k}'',\mathbf{k}) \delta(k-k'') \right].
\end{multline}

\section{Asymmetric scattering for the PD trimers}
The wave function of graphene for the electrons near the K and K' points of Brillouin zone reads as follows:
\begin{equation} \label{WF_cont}
\psi_{\mathbf{k}}(\mathbf{r})=\left(\begin{array}{c}
1 \\ e^{\pm i \theta}
\end{array}\right)\frac{1}{\sqrt{2S}} \exp(i \mathbf{k} \mathbf{r}),
\end{equation}
where $S$ indicates the surface of the sample and $\theta$ is the electron wave vector polar angle. The potential of three--point defects arranged at the corners of a triangle is given by
\begin{equation} \label{PDtrimer1}
V(\mathbf{r})=V_0 S_d \sum_{i=1,2,3} \delta(\mathbf{r}/L-\tilde{\mathbf{r}}_i),
\end{equation}
where $L$ defines the size of the trimer, $\tilde{\mathbf{r}}_1 =(0,1), \tilde{\mathbf{r}}_2 =(\sqrt{3}/2,-1/2), \tilde{\mathbf{r}}_3 =(-\sqrt{3}/2,-1/2)$, and $V_0 S_d$ is the strength of each point defect, namely the product of its depth $V_0$ and surface $S_d$. In this paper tildes indicate normalized and dimensionless values.

The introduced potential of the PD trimer leads to the following matrix element:
\begin{equation} \label{PDtrimer2}
V(\mathbf{k},\mathbf{k}')=\frac{V_0 S_d}{S} \tilde{V}(\tilde{\mathbf{k}},\tilde{\mathbf{k}}'),
\end{equation}
where
\begin{equation} \label{PDtrimer_Vq}
\tilde{V}(\tilde{\mathbf{k}},\tilde{\mathbf{k}}') = \frac{1}{2}\left(1+e^{\pm i(\theta_{\mathbf{k}}-\theta_{\mathbf{k}'})}\right) \sum_{i=1,2,3} \exp(i\tilde{\mathbf{q}}\tilde{\mathbf{r}}_i),
\end{equation}
and $\tilde{\mathbf{q}}=\tilde{\mathbf{k}}-\tilde{\mathbf{k}}'$ is the normalized transferred wave vector ($\tilde{\mathbf{k}}=\mathbf{k}L,\,\tilde{\mathbf{k}}'=\mathbf{k}'L$).

The unique nature of graphene, namely that it is composed of two crystal sublattices, underlies the presence of the multiplier $\frac{1}{2}(1+e^{\pm i(\theta_{\mathbf{k}}-\theta_{\mathbf{k}'})})$ in \eqref{PDtrimer_Vq}, which is absent for electrons in quantum wells. The asymmetric part of the scattering rate \eqref{Wasymm1} after this substitution takes the form
\begin{equation} \label{WasymmPDtrimer}
W^{as}(\mathbf{k},\mathbf{k}')d\mathbf{k}' = d\mathbf{k}'  \frac{V_0^3 S_d^3}{2\pi \hbar^3 v_F^2 S L} \delta(k-k')
\tilde{W}^{as}(\tilde{\mathbf{k}},\tilde{\mathbf{k}}'),
\end{equation}
where
\begin{multline} \label{WasNormalizedDefinition}
\tilde{W}^{as}(\tilde{\mathbf{k}},\tilde{\mathbf{k}}')
= \Im \left[ \tilde{V}(\tilde{\mathbf{k}},\tilde{\mathbf{k}}') \right. \\ \times
\left. \int \tilde{k}''d\tilde{k}''d\theta'' \tilde{V}(\tilde{\mathbf{k}}',\tilde{\mathbf{k}}'')\tilde{V}(\tilde{\mathbf{k}}'',\tilde{\mathbf{k}}) \delta(\tilde{k}-\tilde{k}'') \right]
\end{multline}
can be analytically calculated for the case of the PD trimer.
The general form of the asymmetric part of scattering rate for an arbitrary $\tilde{k}$ provided in the appendix. For $\tilde{k} \ll 1$, the asymmetric part of the scattering rate $\tilde{W}^{as}(\tilde{k},\theta,\phi)$ can be written as
\begin{multline} \label{I_PD_analytic}
\tilde{W}^{as}(\tilde{k},\theta,\phi)=\frac{3\pi\tilde{k}^3}{2}
\frac{(1+\cos(\phi))}{4}\\
 \times \cos\left(\frac{3\phi}{2}+3\theta\right) \left( 3\sin\left(\frac{\phi}{2}\right)+2\sin\left(\frac{3\phi}{2}\right) \right).
\end{multline}

The factor $\frac{1}{4}(1+\cos(\phi))$ stems from the spinor structure of the electronic wave function of electrons in graphene.

\section{Asymmetric scattering for large triangular clusters}

\subsection{Introducing the potential}

For large clusters with $kL \gg 1$ the scattering cross section itself and the asymmetry of scattering as well are substantial only for scattering angles $\phi \ll 1$, and their magnitudes tend to 0 for large angles, which allows to show that the spinor structure of the graphene wave function containing the multiplier $e^{\pm i \theta}$ does not affect the skew scattering features of large clusters. Therefore, in this section we assume that $\psi_{\mathbf{k}}(\mathbf{r})=\frac{1}{\sqrt{S}} \exp(i \mathbf{k} \mathbf{r})$.

\begin{figure}[!ht]
\includegraphics[width=1.0\linewidth]{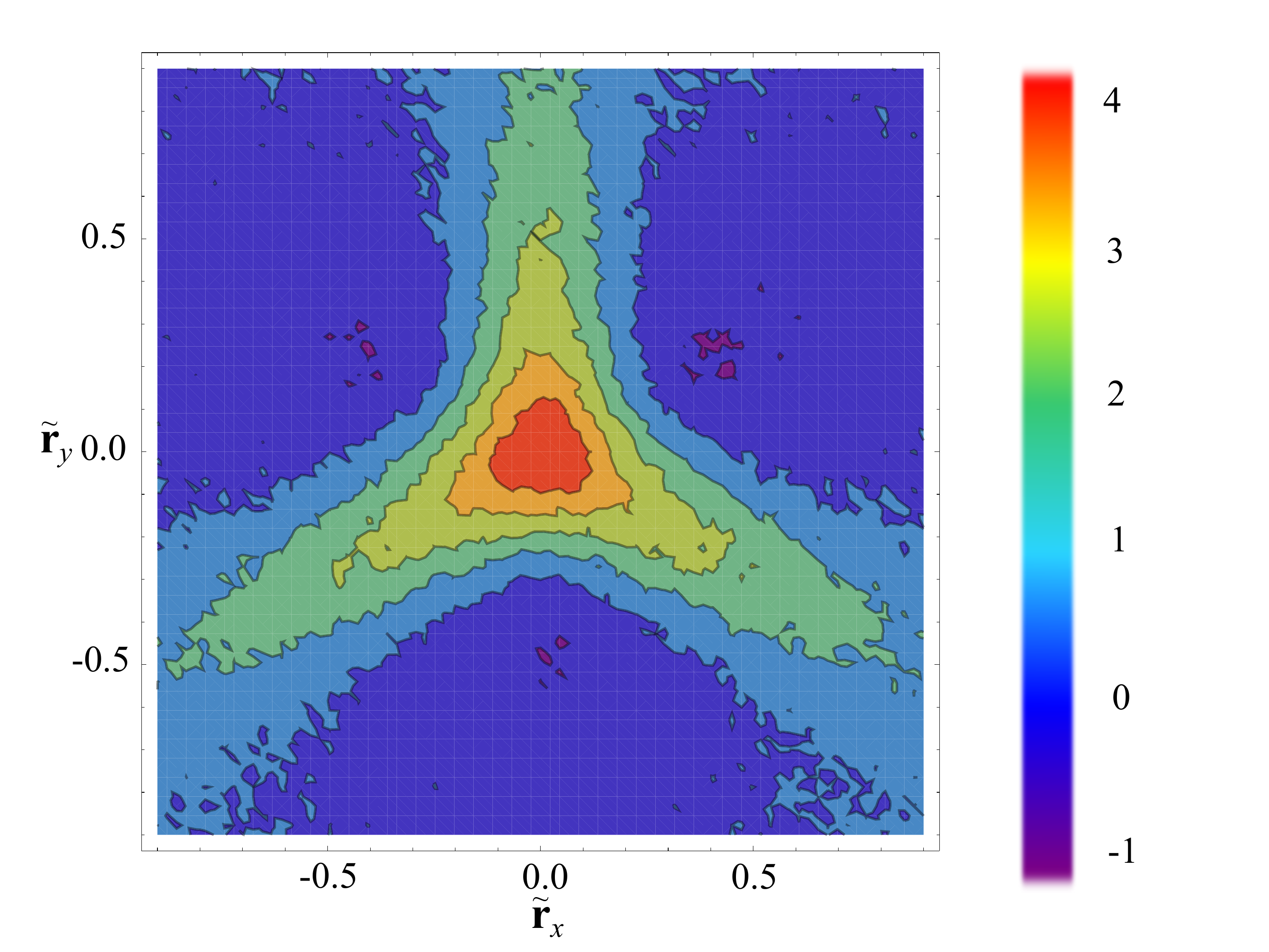}
\caption{The scattering potential $\tilde{V}(\tilde{\mathbf{r}})$ in real space calculated based on the Fourier image \eqref{img_potential}. The torn shape of the equipotential curves is due to the Monte Carlo integration method.}
\label{fig_RealSpace}
\end{figure}

It is natural to introduce the potential of the scatterer in the following form:
\begin{equation} \label{ME1}
V(\mathbf{r})=V_0 \tilde{V}(\mathbf{r}/L),
\end{equation}
where $V_0$ is the binding energy of the impurities of which the triangular cluster is composed, for instance adatoms, functional groups, vacancies or the relief of the substrate. $L$ is the characteristic size of the cluster, and $\tilde{V}(\tilde{\mathbf{r}})$ gives the shape of the potential of the cluster, which differs from zero for $|\tilde{\mathbf{r}}|$ on the order of 1. The matrix element is consequently written as
\begin{equation}\label{ME2}
V(\mathbf{k},\mathbf{k}')=\frac{V_0L^2}{S}\tilde{V}(\tilde{\mathbf{q}}),
\end{equation}
where
\begin{equation}\label{ME3}
\tilde{V}(\tilde{\mathbf{q}})=\int d\tilde{\mathbf{r}} \tilde{V}(\tilde{\mathbf{r}}) \exp(i\, \tilde{\mathbf{q}} \tilde{\mathbf{r}}).
\end{equation}
The real and the imaginary parts of the scatterer's potential fourier image are proposed as Gaussian functions modulated by angle, which in polar coordinates, $\tilde{\mathbf{q}}=(\tilde{q},\alpha)$, can be expressed as follows:
\begin{subequations}
\label{img_potential}
\begin{align}
\Re[V(\tilde{\mathbf{q}})]=\tilde{V}_0 \exp(-4\tilde{q}^2(2-\cos (6\alpha))),\\
\Im[V(\tilde{\mathbf{q}})]=\tilde{V}_0 \exp(-2\tilde{q}^2(2+\cos (6\alpha))) \sin (3\alpha).
\end{align}
\end{subequations}
Assuming $\tilde{V}_0 = 0.026$ provides the correct normalization of the potential $\int \tilde{V}(\tilde{\mathbf{r}})d\tilde{\mathbf{r}} = 1$. Fig. \ref{fig_RealSpace} shows that the potential $\tilde{V}(\tilde{\mathbf{r}})$ in real space, built based on its Fourier image \eqref{img_potential}, describes the triangulaer potential with good accuracy. Integral calculation is performed via the Monte Carlo method, which minimizes the artifacts of the numerical integrating procedure. The meaningful and reproducible result of the numerical integration confirms the acceptable accuracy of the calculations.

Thus, the asymmetric part of the scattering rate reads as
\begin{equation} \label{WasymmFinal}
W^{as}(\mathbf{k},\mathbf{k}')d\mathbf{k}' = d\mathbf{k}'  \frac{V_0^3L^5}{2\pi \hbar^3 v_F^2 S} \delta(k-k')
\tilde{W}^{as}(\tilde{\mathbf{k}},\tilde{\mathbf{k}}'),
\end{equation}
where we numerically calculate the normalized dimensionless asymmetry of scattering $\tilde{W}^{as}(\tilde{\mathbf{k}},\tilde{\mathbf{k}}')$, defined straightforwardly as in \eqref{WasNormalizedDefinition}. This integral is also treated by the Monte Carlo method. The Dirac delta function is taken as the narrow Lorentz peak: $\delta(x)=\frac{a}{\pi}(a^2+x^2)^{-1}$, where $a \ll 1$.

\subsection{Asymmetric scattering for large clusters}

The asymmetric part of the scattering rate obeys the following symmetry relation: $\tilde{W}^{as}(\theta,\phi)=\tilde{W}^{as}(\theta+\frac{2\pi}{3},\phi)$. For small scattering angles $\phi$, an additional symmetry rule is met: $\tilde{W}^{as}(\theta,\phi)=-\tilde{W}^{as}(\theta,-\phi)$.

The following approximation of the scattering asymmetry can be written for all values of the normalized electron wave vector $\tilde{\mathbf{k}}$:

\begin{equation} \label{I_large_analytic}
\tilde{W}^{as}(\tilde{k},\theta,\phi)=\begin{cases} A \cos (3\theta) (4.25-\tilde{k} |\phi|) \text{sign}(\phi),\\
|\phi|<4.25/\tilde{k};\\
0, |\phi|>4.25/\tilde{k},
\end{cases}
\end{equation}
where $A=0.000058$, which leads to the following integral relations: $\int_{-\pi}^{+\pi} \tilde{W}^{as}(\tilde{k},\theta,\phi) d\phi = 0$ and $\int_{-\pi}^{+\pi} \tilde{W}^{as}(\tilde{k},\theta,\phi) \phi d\phi \approx 9.3\cdot 10^{-6}\cos{3\theta}$. The transferred wave vector has a value $k\phi$.

\begin{figure}
\includegraphics[width=1.0\linewidth]{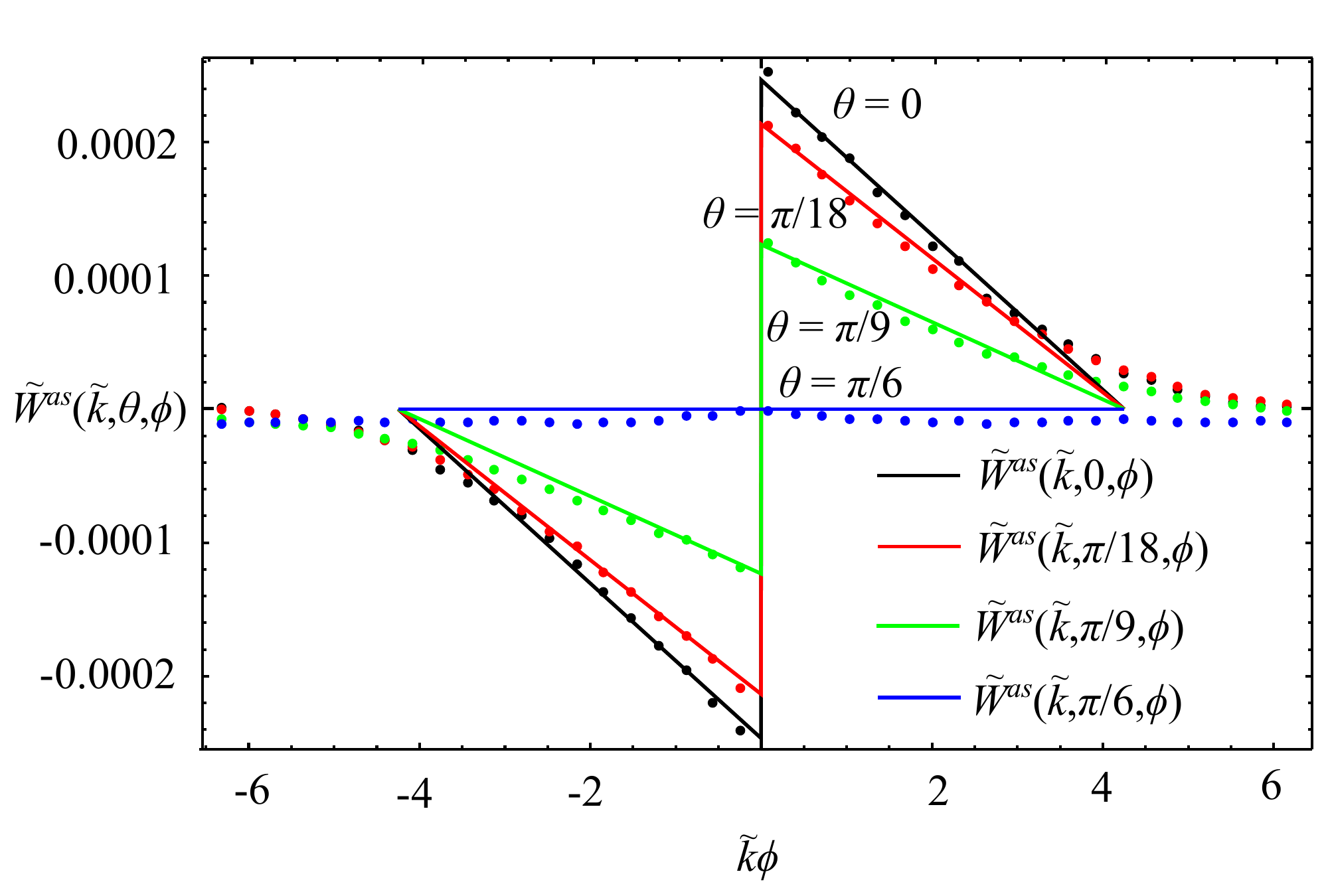}
\caption{Angular dependence of the asymmetric part of the scattering rate  the large clusters. Dots represent the calculated dependence of the normalized asymmetry scattering $\tilde{W}^{as}$ on the transferred wave vector $\tilde{k}\phi$ for different values of $\theta$ and $\tilde{k}=4$. Black indicates $\theta=0$, red indicates $\theta=\pi/18$, green indicates $\theta=\pi/9$ and blue indicates $\theta=\pi/6$. The solid lines with a similar color scheme are plots of \eqref{I_large_analytic} for corresponding values of $\theta$.}
\label{fig_WonPhi}
\end{figure}

The asymmetric scattering operator reads as
\begin{equation} \label{operator_asymm}
\hat{A}(f(\mathbf{k}'))=\frac{1}{4 \pi^2}\int_{BZ} d\mathbf{k}(W^{as}(\mathbf{k},\mathbf{k}')\left[f(\mathbf{k})+f(\mathbf{k}')\right],
\end{equation}
After combining $W^{as}$ from \eqref{WasymmFinal} and \eqref{I_large_analytic}, substituting $f(\theta)+f(\theta + \phi) \approx 2f(\theta) + \phi \partial f(\theta) /\partial \theta$ and integrating over $k$, one obtains
\begin{multline} \label{f2_theta}
\hat{A}(f(\mathbf{k})) = \frac{V_0^3 k L^5}{8\pi^3 \hbar^3 v_F^2 S}\\ \times
\int_{-\pi}^{+\pi} d\phi \tilde{W}^{as}(\theta-\phi,\phi) \frac{\partial f(\theta)}{\partial \theta} \phi.
\end{multline}
The integral over $\phi$ that contains $f(\theta)$ vanishes and therefore, only the angular derivative of the distribution function $\phi \partial f/\partial \theta$ gives rise to the effect of asymmetric scattering. Integrating over $\phi$ one yields
\begin{equation} \label{f2_theta}
\hat{A}(f(k,\theta)) =9.3\cdot10^{-7}  \frac{V_0^3 k L^5 }{8\pi^3 \hbar^3 v_F^2 S} \cos(3\theta) \frac{\partial f(k,\theta)}{\partial \theta}.
\end{equation}

\section{Asymmetric scattering in classic limit}

In this section we will describe the cross section for the elastic scattering of a point particle on a hard triangle oriented as shown on Fig. \ref{fig_Geometry}. The elementary geometry leads to the following form of the rate of the scattering cross section for the direction $\theta$ of the initial electron wave vector $\mathbf{k}$ lying in the range between 0 and $\frac{2}{3}\pi$:
\begin{equation} \label{f2_theta}
W(\theta,\phi)= \frac{v_F L}{S}\frac{\delta(k-k')}{k}G(\theta,\phi),
\end{equation}
where
\begin{equation} \label{f2_theta}
G(\theta,\phi)=\begin{cases}
\delta(\phi-(\frac{2}{3}\pi-2\theta)) \sin(\pi/3 - \theta) \\ + \delta(\phi+2\theta) \sin(\theta), 0 < \theta < \frac{2}{3}\pi; \\ \delta(\phi+2\theta)\sin(\theta), \frac{2}{3}\pi < \theta < \pi/2 ;\\
\delta(\phi-(2\pi-2\theta))\sin(\theta), \pi/2 < \theta < \frac{4}{3}\pi.
\end{cases}
\end{equation}
$L$ in the equation above represents the length of one side of the triangle. The dependence of the cross section on the angle $\theta$ has a period $\frac{2}{3}\pi$, and consequently, it can be extended to the entire range of angle $\phi$ from 0 to $2\pi$. The asymmetric part of the scattering rate $\tilde{W}^{as}(\theta,\phi)$ can be written as $\tilde{W}^{as}(\theta,\phi)=G(\theta,\phi)-G(\theta+\phi,-\phi)$, and finally
\begin{equation} \label{f2_theta}
W^{as}(\theta,\phi)= \frac{v_F L}{S}\frac{\delta(k-k')}{k}\tilde{W}^{as}(\theta,\phi).
\end{equation}

\section{Results and discussion}

\subsection{Kinetics}

In the case of weak asymmetry when $W^{as}(\mathbf{k},\mathbf{k}') \ll W^{symm}(\mathbf{k},\mathbf{k}')$ considered in this paper, the ratchet effect will appear in the correction to the electron distribution function linear to the asymmetric scattering rate and quadratic to the electric field magnitude, which will require three iterations of the Boltzmann kinetic equation solution. Moreover the two contributions of the same magnitude and similar structure will give raise to the effect. We write the electric field as $\mathbf{E}(t)=\mathbf{E} e^{i\omega t}+ C.C.$
The methods for obtaining the iterative solution to the Boltzmann kinetic equation for nonlinear transport in graphene are well developed \cite{chep3,Glazov,PhysRevB.86.115301}, and we generally adhere to them. The first contribution to the correction for the carrier distribution function is written as

\begin{equation}
f^{I}=\tau\frac{e\mathbf{E}^{*}}{\hbar} \frac{\partial }{\partial \mathbf{k}}\frac{\tau}{1-i \omega \tau} \hat{A} \frac{\tau}{1-i \omega \tau} \frac{e\mathbf{E}}{\hbar} \frac{\partial }{\partial \mathbf{k}} f^{(0)}(k,\theta),
\end{equation}
and the second one is written as
\begin{equation}
f^{II}=\tau \hat{A} \tau \frac{e\mathbf{E}^{*}}{\hbar} \frac{\partial }{\partial \mathbf{k}}\frac{\tau}{1-i \omega \tau} \frac{e\mathbf{E}}{\hbar} \frac{\partial }{\partial \mathbf{k}} f^{(0)}(k,\theta).
\end{equation}
$f^{(0)}$ in the equations above is the Fermi-Dirac equilibrium distribution function, and $\tau$ is the relaxation time for the corresponding angular harmonic of the electron distribution function correction. The ratchet current reads as follows:
\begin{equation} \label{f2_theta}
\left(\begin{array}{c}
j_x \\ j_y
\end{array}\right)
=\frac{e v_F}{\pi^2}\int kdk d\theta \left(\begin{array}{c}
\cos(\theta) \\ \sin(\theta)
\end{array}\right) (f^{I}+f^{II}).
\end{equation}

We assume that the symmetric part of the scattering rate is defined by the intrinsic mechanisms of the sample and take into account the two common mechanisms of the intrinsic conductivity of graphene: scattering on Coulomb defects and on short-range defects. The difference will appear in the dependencies of the ratchet current on the electric field frequency.

Scattering on Coulomb impurities leads to the following dependencies of the relaxation times of the first (proportional to $\sin \theta$ and to $\cos \theta$) and of the second (proportional to $\sin 2\theta$ and to $\cos 2\theta$) angular harmonics of the distribution function correction on the absolute value of the electron wave vector \cite{RevModPhys.81.109}:
\begin{equation} \label{CoulombRelax}
\tau_1(k)=\tau_{tr} \frac{k}{k_F},\quad \tau_2(k)=3\tau_1,
\end{equation}

For short-range impurities, the dependencies of the relaxation times on the wave vector are given by
\begin{equation} \label{ShortRelax}
\tau_1(k)=\tau_{tr} \frac{k_F}{k},\quad \tau_2(k)=\tau_1/2.
\end{equation}

$\tau_{tr}$ in the equations above is the transport relaxation time of a given graphene sample.

\subsection{Analytical expressions for the ratchet current}
TABLE I shows the analytical expressions for the ratchet current for the different types of triangular scatterers considered above. The designation
\begin{equation*} \label{f2_theta}
\tau_{n, \omega} = \frac{\tau_n(k)}{1-i \omega \tau_n(k)},
\end{equation*}
appears in the column corresponding to the arbitrary intrinsic scattering mechanism. Therefore, the real part of the given coefficients $\chi$ is assumed to derive the physical value of the ratchet current.  $\tau'$ implies differentiation by the magnitude of the electron wave vector and $n$ represents the concentration of trigonal clusters.

\begin{table*}[!ht] \label{Table_Anal}
\caption{Analytical relations for ratchet current for the considered types of triangular clusters (de facto coefficient $\chi$ from relation \eqref{symmetry}). The relations in the left column contains dependencies of the ratchet current on the relaxation times of the first and the second angular harmonics for the arbitrary intrinsic conductivity of the sample. In the last two columns the relations \eqref{CoulombRelax} and \eqref{ShortRelax} are substituted for the scattering on Coulomb and short-range defects respectively.}

\begin{tabular}{ c  c  c  c }
\hline
\hline
\multirow{2}{*}{Scatterer}  & \multicolumn{3}{c}{Intrinsic scattering mechanism}\\
\cline{2-4}
 & Arbitrary mechanism & Coulomb defects & short-range defects \\
\hline
PD trimer
&
$\begin{array}{l} \chi_{tr}= \tau_{1,\omega} (6\tau_1(\tau_{2}-\tau_{2,\omega}) \\+k_F(\tau_{1}'(\tau_{2}+6\tau_{2,\omega})+\tau_{1}\tau_{2}')) \\ \times
\dfrac{V_0^3 S_d^3 k_F^2}{64\pi^2 \hbar^5 v_F} (k_FL)^3 e^3n. \end{array}$
&
$\chi_{tr,C}= \dfrac{21\tau_{tr}^3}{256(1+\omega^2 \tau_{tr}^2)}
\dfrac{V_0^3 S_d^3 k_F^2 }{\pi^2 \hbar^5 v_F} (k_FL)^3 e^3n$
&
$\begin{array}{l}\chi_{tr,s}= \dfrac{(-28+29\omega^2 \tau_{tr}^2) \tau_{tr}^3}{512(1+\omega^2 \tau_{tr}^2)(1+4\omega^2 \tau_{tr}^2)}\\ \times
 \dfrac{V_0^3 S_d^3 k_F^2}{\pi^2 \hbar^5 v_F} (k_FL)^3 e^3n\end{array}$
\bigskip
\\
\shortstack{Large \\ clusters}&
$\begin{array}{l l} \chi_l= \tau_{1,\omega} (-\tau_{2,\omega} \tau_1 + 6 \tau_1 \tau_2  \\ + k_F (\tau_{2,\omega} \tau_1' + 2 \tau_2 \tau_1' + 2 \tau_1 \tau_2') ) \\
\times \dfrac{2.9\cdot10^{-8}V_0^3 k_F L^5}{\pi^4 \hbar^5 v_F} e^3 n \end{array}$
&
$\chi_{l,C}= \dfrac{6.9\cdot10^{-7} \tau_{tr}^3}{1+\omega^2 \tau_{tr}^2}
 \dfrac{V_0^3 k_F L^5}{\pi^4 \hbar^5 v_F} e^3 n$
&
$\chi_{l,s}= \dfrac{1.2\cdot10^{-7} \tau_{tr}^3}{1+ 4 \omega^2 \tau_{tr}^2}
 \dfrac{V_0^3 k_F L^5}{\pi^4 \hbar^5 v_F} e^3 n$
\bigskip
\\
\shortstack{Classic \\ scattering}
&
$\begin{array}{l} \chi_{cl}= \tau_{1,\omega} (\tau_1 (\tau_{2,\omega}+2\tau_2) \\
+k_F (\tau_1'\tau_2 - \tau_{2,\omega}\tau_1' + \tau_1 \tau_2')) \\ \times
\dfrac{3v_F^2 L}{32 \pi^3 \hbar^2} e^3n \end{array}$
&
$\chi_{cl,C}=\dfrac{9\tau_{tr}^3}{32(1+\omega^2 \tau_{tr}^2)}
 \dfrac{v_F^2L}{\pi^3 \hbar^2} e^3n$
&
$\chi_{cl,s}=\dfrac{3(4-\omega^2 \tau_{tr}^2) \tau_{tr}^3}{64(4+\omega^2 \tau_{tr}^2)(1+\omega^2 \tau_{tr}^2)}
 \dfrac{v_F^2L}{\pi^3 \hbar^2} e^3 n$ \medskip \\
\hline
\hline
\end{tabular}
\end{table*}

Substituting the relaxation times for scattering on the Coulomb defects \eqref{CoulombRelax} and on the short-range defects \eqref{ShortRelax} yields the relations in the corresponding columns of TABLE I.

In the results presented in TABLE I, we do not explicitly use the concentration of carriers $n_s=p_F^2/(\pi\hbar^2)$, where the Fermi momentum $p_F=\hbar k_F$. If one uses the carrier concentration, the Planck constant will vanish in the solution for the classical scattering, and the solution will remain a cubic function for quantum mechanical scattering.

In contrast with ref.\cite{PhysRevB.86.115301}, which studied ratchet effects in noncentrosymmetric 1D periodic patterns on graphene and ref. \cite{chep3}, which studied skew scattering on semidisks (both studied systems with $C_{2v}$ symmetry), $C_{3v}$ symmetry allows the photocurrent to appear only as a response to the polarized light, and no current is driven by the unpolarized radiation. In ref. \cite{PhysRevB.86.115301} and ref. \cite{chep3}, the $X$ and $Y$ components of the ratchet current are the superposition of both the polarized and unpolarized radiation contributions. A similar situation occurs for for the photon drag current in graphene and the edge photocurrents \cite{Glazov,PhysRevLett.107.276601}. Therefore, the contribution from skew scattering on the trigonal clusters can be separated from other contributions by the polarization dependence. So in graphene with trigonal clusters, the $Y$ and $X$ components of the ratchet current indicate the degree of linear polarization in the $X$ and $Y$ axes and the axes rotated by 45$^{\circ}$, respectively.

The frequency dependencies of the ratchet currents are generally similar to those reported in refs. \cite{PhysRevB.86.115301, chep3}, namely rational functions of $(\omega \tau_{tr})^2$. The behavior of the ratchet current $(\omega \tau_{tr})^{-2}$ corresponding to Drude absorption describes large frequencies for all types scatterers considered.

In ref. \cite{PhysRevB.86.115301}, the transverse ratchet effect, i.e., the situation in which the ratchet current and driving force are perpendicular, is impossible for graphene with Coulomb defects. This paper shows that skew scattering on trigonal clusters makes the transverse ratchet effect possible for all actual mechanisms of intrinsic conductivity.

It is important to underscore the difference between classic and quantum mechanical scattering on a macroscopic triangular cluster. The difference is the strength of the potential. In the case of classic scattering, the potential barrier is assumed to be much higher than the Fermi level in the system. It is important to reduce the effects of the Klein tunneling \cite{Kats}. If the potential is weak with respect to the Fermi energy, the Born approximation is acceptable, and one should use the theory for large clusters. Quantum well are more preferable to barriers in maintaining the accuracy of the Born approximation.

The value of $\tilde{k}=kL$, where $L$ is the length of one side of a triangular cluster, defines the quantum mechanical regime of electron scattering \cite{PhysRevB.79.195426}. For clusters that are large compared with the Fermi wavelength ($\tilde{k} \gg 1$), the first Born approximation  does not reproduce the side lobes of the scattering cross--section, and therefore, the applicability of the approximation should be ascertained. However the first Born approximation describes forward scattering as being tolerable.

The developed approach appears to be unsuitable for treating deep quantum wells and high barriers, but for $V_0 \ll \varepsilon_F$, one can expect acceptable results. It is important to note that threefold quantum dots exhibit intriguing properties of localized electronic states \cite{1367-2630-10-10-103015,PhysRevB.56.12147}.

The suggested ratchet effect can be detected in the following experiments. The first experiment implies detection of photogalvanic current due to irradiation of the graphene sheet by linearly polarized electromagnetic waves. The highest radiation frequency that do not suppress the amplitude of the effect is determined by the transport relaxation time of carriers in graphene sample $\tau_{tr}$ and calculation for the adopted parameters of the sample yields 16\,THz. The second experiment is detection of the transverse ratchet effect in the Hall effect geometry, and moreover the AC to DC conversion is expected to be observed in this case.

\subsection{Numerical estimations of the ratchet current values}

In this section, we provide numerical estimations of the obtained ratchet current for the actual and reachable parameters of the sample. All obtained expressions for the ratchet current contain both the relaxation time and Fermi level in the sample. From these two quantities, the electrical conductivity of the sample can be directly derived as follows:
\begin{equation*} \label{AppB1}
\sigma=\frac{e^2 v_F k_F}{\pi \hbar} \tau_1(k_F).
\end{equation*}
As mentioned above, the carrier concentration $n_s$ is related to the Fermi wave vector via $k_F=\sqrt{\pi n_s}$ (see also the very useful TABLE II in the review \cite{RevModPhys.83.407}).

To numerically estimate the Fermi vector and relaxation time, the parameters of the sample described in the classic paper studying graphene conductivity \cite{PhysRevLett.98.186806} were adopted. The authors provide an explanation of the conductivity of the studied sample by scattering on Coulomb defects. FIG. 5 from ref. \cite{PhysRevLett.98.186806} allows for the estimation $\tau_{tr} \approx 6.3\, 10^{-14}\text{s}$ for $k_F \approx 3.96\, 10^{6} \text{cm}^{-1}$, which corresponds to $\varepsilon_F \approx 0.24 \text{eV}$ and $\lambda_F\sim 15 \text{nm}$. The mean free path $\tau_{tr} v_F$ is consequently 63\,nm.

Finally, it is important to estimate the magnitude of the actual electric fields. A radiation intensity 1 W/cm$^2$ corresponds to an electric field strength of approximately 20 V/cm. By the 1980s, a power several orders of magnitude higher had already been reached by semiconductor lasers \cite{RevModPhys.73.767}. A radiation intensity of only $10^{14-15}$\,W/cm$^2$ yields an electric field of the same magnitude as the Coulomb field inside atoms\cite{RevModPhys.84.1177}.

TABLE II presents the numerical estimations of the ratchet current in a graphene sample with the parameters adopted from \cite{PhysRevLett.98.186806}.

\begin{table*}[!ht] \label{Table_Num}
\caption{Numerical estimations of the ratchet current. The parameters of the sample with conductivity defined by scattering on the Coulomb impurities are: $\varepsilon_{F}=0.24$\,eV, $\tau_{tr} \approx 6.3\, 10^{-14}\text{s}$. $\tau_c$ is the transport relaxation time for scattering on the triangular clusters with given parameters only. The concentration of triangular scatterers $n$ is assumed to be $1 \mu m^{-1}$. The values of the ratchet current $j$ are presented in the last two columns per the square of the electric field magnitude and per the intensity of radiation.}
\begin{tabular}{ l c c c c c }
\hline
\hline
Scatterer &
Scatterer parameters &
$k_FL$&
$\tau_c$, s&
$j$, $\text{pA}\,\text{cm} \text{V}^{-2}$ &
$j$, $\text{pA}\,\text{cm} \text{W}^{-1}$ \\
\hline
PD trimer &
$V_0S_d=0.1$\,eV\,nm$^2$, $L=1$\,nm&
0.4 &
3.3 $10^{-8}$ & $3.4\cdot 10 ^{-6}$ & $1.4\cdot 10 ^{-3}$ \\
Large cluster&
$V_0=0.1$\,eV, $L=10$\,nm &
4 & $1.5 \cdot 10^{-10}$ & $2.6 \cdot 10^{-6}$ & $1.0 \cdot 10^{-3}$ \\
Classic scattering &
$L=100$\,nm&
40&
$10^{-11}$ & $28$ & $1.1 \cdot 10 ^{4}$ \\
\hline
\hline
\end{tabular}
\end{table*}

All results are presented for a steady external electric field ($\omega=0$) and the actual depth and size of the scatterers (TABLE II). The concentration of trigonal clusters is assumed to be $n = 1\, \mu m^{-1}$. It is trivial to extend the results to arbitrary electric field and scatterers parameters.

In addition, we have ascertained that the contribution to the transport relaxation time $\tau_c$ from the trigonal clusters is much smaller than the contribution from the intrinsic defects. Quantum mechanical consideration of scattering on the short-range defects gives
\begin{equation*} \label{V0_cont}
\tau_c \sim \frac{\hbar^2 v_F}{V_0^2 S_{d}^2 n k_F},
\end{equation*}
where $V_0$ is the defect depth and $S_d$ provides the estimate of the defect area. For large clusters, $S_d \approx L^2$. The extra multiplier $(k_FL)^2$ will appear for the large clusters due to the strong forward scattering. In the classical, consideration one estimates $\tau_c \sim (n L v_F)^{-1}$.

It can be observed that the ratchet current for the small PD trimer is comparable to the current for a large cluster with a similar potential depth, although the potential strength of the PD trimer, which appears to be cubed , is $10^2$ smaller. The first reason whys this figure is obtained is that the large cluster can transfer only a small wave vector on the order of $L^{-1}$ to electrons, which suppresses both backscattering and skew scattering. The second reason is that the shape of the solid triangle is much more similar to that of the isotropic disk scatterer than that of the PD trimer.

Scattering in the classic regime exhibits a stronger effect than scattering in the QM regime with a weak potential for triangles of the same size.

The values of the transport relaxation times $\tau_c$ on all types of threefold clusters are much smaller than the characteristic intrinsic transport relaxation times of graphene samples. Therefore, the developed theory will remain applicable for concentrations several orders of magnitude greater than those of the PD trimers. An increase in the large triangular cluster concentration is nearly impossible due to geometrical constraints.

The current technologies used for manufacturing \cite{GeimNat,Dideykin2011105} and treating graphene via laser \cite{0957-4484-22-47-475303,C2NR30790A} and chemical \cite{doi:10.1021/ja4042077} patterning and ion etching \cite{NatEt1,doi:10.1021/ja402224h} provides the greatest possibilities for creating ratchet devices with the proposed geometrical structure. Another way to obtain threefold scatterers in graphene is to grow epitaxial graphene films on substrates with threefold islands or to place graphene sheets on substrates patterned with triangles. The obtained values of the ratchet photocurrents demonstrate the potential of creating graphene-based polarization sensitive detectors of THz and microwave radiation.

\appendix
\section{Asymmetric scattering on the PD trimer for arbitrary electron wave vector}

Substituting the PD trimer matrix element \eqref{PDtrimer_Vq} into the general formula for the asymmetric scattering rate \eqref{Wasymm1} leads integration over the wave vector absolute value and to the angular integral. Integrating over $\tilde{k}''$ and using the energy conservation law $\delta(\tilde{k}-\tilde{k}')$ will yield $\tilde{k}=\tilde{k}'=\tilde{k}''$. The unclosed angular integral will contain the exponent of sine and cosine functions of the angles between the wave vectors $\tilde{\mathbf{k}},\tilde{\mathbf{k}}',\tilde{\mathbf{k}}''$ and the PD radius vectors $\tilde{\mathbf{r}}_1, \tilde{\mathbf{r}}_2, \tilde{\mathbf{r}}_3$. The form of the integral is the sum of the expressions that match the left-hand side of the equation
\begin{multline} \label{App1}
\int_{-\pi}^{\pi} \exp(i\tilde{k}A\cos(\theta'')+i\tilde{k}B\cos(\theta'')+i\tilde{k}C) d\theta''=\\
2\pi\exp(i\tilde{k}C)J_0\left(\tilde{k}\sqrt{A^2+B^2}\right),
\end{multline}
and of the equation
\begin{multline} \label{App1}
\int_{-\pi}^{\pi} \exp(i\tilde{k}A\cos(\theta'')+i\tilde{k}B\cos(\theta'')+i\tilde{k}C) \exp(\pm i\theta'') d\theta''=\\
-2i\pi\exp(i\tilde{k}C)\\ \times \left(J_1\left(-\tilde{k}\sqrt{A^2+B^2}\right) \pm \frac{2 \sin(\tilde{k}\sqrt{A^2+B^2})}{\tilde{k}\sqrt{A^2+B^2}} \right),
\end{multline}
where the parameters $A$ and $B$ depend only on the polar angles $\tilde{\mathbf{r}}_1, \tilde{\mathbf{r}}_2, \tilde{\mathbf{r}}_3$, and the coefficient $C$ contains the sine and cosine function of $\theta$ and $\theta'=\theta+\phi$.

Further simplifications lead to expression $\eqref{I_PD_analytic}$ for a dimensionless asymmetric scattering rate.

\acknowledgements

I am gratefully indebted to L.E. Golub and M.M. Glazov for invaluable support and stimulating discussions; to E.D. Eidelman and A.Ya.Vul’ for their support; and to O.I. Utiosov for discussions on scattering theory. This work was supported by the Dynasty Foundation, by Ioffe Physical-Technical Institute project "Physical-chemical principles of new functionalized materials based on carbon nanostructures" and by N.M. Dybkov, a kind of Maecenas for me.

%

\end{document}